\documentclass[twocolumn,12pt,cites,graphicx]{article}
\usepackage{epsfig}
\usepackage{subfigure}

\title{Bicontinuous emulsions stabilized solely by colloidal particles}
\author{E.~M.~Herzig  $^{\dagger \ddagger}$ \and K.~A.~White$^{\dagger \ddagger}$ \and A.~B.~Schofield$^{\dagger}$ \and W.~C.~K.~Poon$^{\dagger \ddagger}$ \and  P.~S.~Clegg\thanks{email: pclegg@ph.ed.ac.uk} \thanks{SUPA, School of Physics, University of Edinburgh, Mayfield Road, Edinburgh, EH9 3JZ, UK} \thanks{COSMIC, University of Edinburgh, Mayfield Road, Edinburgh, EH9 3JZ, UK}}
\begin{document}

\maketitle
\renewcommand{\thefootnote}{\fnsymbol{footnote}}

\noindent \textbf{Recent large-scale computer simulations suggest that it may be possible to create a new class of soft solids, called `bijels', by stabilizing and arresting the bicontinuous interface in a binary liquid demixing via spinodal decomposition using particles that are neutrally wetted by both liquids~\cite{stratford05}. The interfacial layer of particles is expected to be semi-permeable~\cite{dinsmore02}, hence, if realised, these new materials would have many potential applications, e.g. as micro-reaction media. However, the creation of bijels in the laboratory faces serious obstacles~\cite{uslangmuir}. In general, fast quench rates are necessary to bypass nucleation, so that only samples with limited thickness can be produced, which destroys the three-dimensionality of the putative bicontinuous network. Moreover, even a small degree of unequal wettability of the particles by the two liquids can lead to ill-characterised, `lumpy' interfacial layers and therefore irreproducible material properties. Here we report a reproducible protocol for creating three-dimensional samples of bijel in which the interfaces are stabilized by essentially a single layer of particles. We demonstrate how to tune the mean interfacial separation in these bijels, and show that mechanically, they indeed behave as soft solids.}

Thermally induced demixing provides a route to creating varied arrangements of fluid-fluid interfaces. On a mean-field level two kinetic pathways to demixing exist: nulceation where droplets of the minority phase coarsen if they exceed a threshold size and spinodal decomposition where the mixed phase becomes unstable and separates into a bicontinuous arrangement of domains~\cite{debenedetti}. Because the miscible region is adjacent to the nucleation region for all compositions apart from near the critical point, spinodal decomposition can be accessed either by quenching very fast through the nucleation region or directly by tuning the system close to the critical composition. For spinodal decomposition a permanent, bicontinuous arrangement of domains will form if the interface can be held in place. 

Here we make use of critical 2,6-lutidine-water mixtures which demix on warming~\cite{grattoni93} and we pin the domains using a jammed monolayer of colloidal particles which are trapped at the interface. Trapping occurs because the colloids reduce the shared area betweeen immiscible fluids an effect which is most pronounced for a fluid-fluid-colloid wetting angle of 90$^{\circ}$ (neutral wetting) and high interfacial tension~\cite{aveyard03, binksbook}. If the wetting angle differs greatly from 90$^{\circ}$ the colloids will induce a preferred curvature in the interface~\cite{aveyard03, binksbook} and this may lead to an excess of particles in one of the phases. We use fluorescent silica colloids with carefully tuned surface chemistry (see Methods). While the choice of a critical composition mixture is a general route to three-dimensional structure formation the colloid wetting properties must be adjusted for each specific system. Our confocal microscopy data shows how these bicontinuous interfacially jammed emulsion gels~\cite{stratford05, patent} (bijels) can be formed and tuned reproducibly.

\begin{figure}[htp]
\centering
\includegraphics[width=0.5\textwidth]{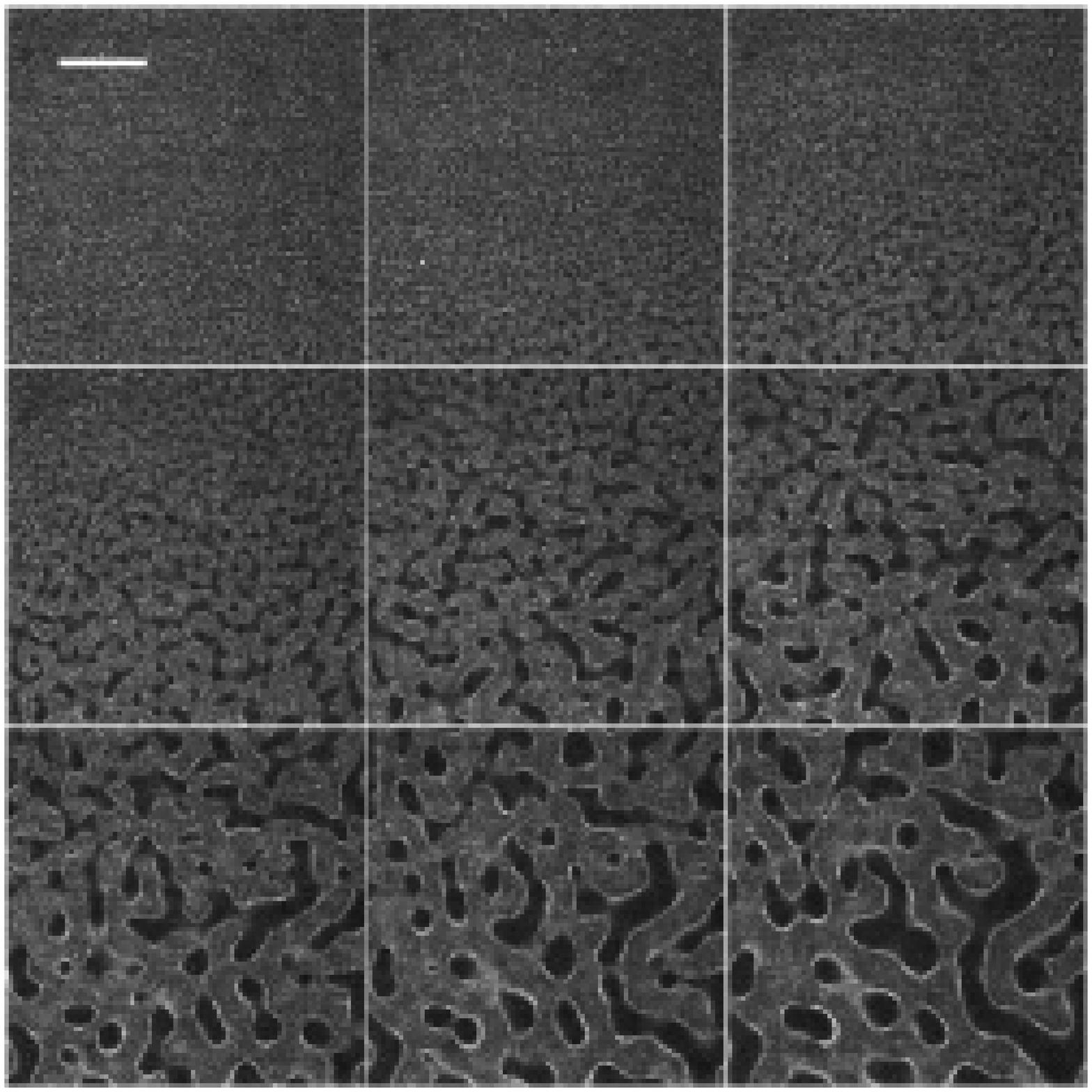}
\caption{{\bf Formation of bijel via phase separation.} Time series of fluorescence confocal microscopy images of a 2,6-lutidine-water sample at critical composition with $\Phi_{v}$=2\% particles, slowly quenched from 33.5$^\circ$C to 35.3$^\circ$C. Only images around the separation are shown. $\Delta$t between images is 0.7 seconds. Particles appear white while liquids appear dark; the difference in the shade of grey for the two domains indicates that the lighter phase contains a substantial population of residual particles (scale bar 100~$\mu$m). The separation via spinodal decomposition is clearly visible.}\label{fig:spindec}
\end{figure}

First we show that nucleation can be bypassed by choosing mixtures with critical composition. Fig.~\ref{fig:spindec} is a time series of confocal slices of a critical composition 2,6-lutidine-water sample containing 2\% volume of silica colloids as it is warmed quasi-statically through the critical point (T$_c$=34.1$^{\circ}$C). 
The images are rastered from top to bottom: while a frame begins every 0.7~sec, time also progresses during the scanning of each frame. The characteristic spinodal pattern is clearly visible with the domain size growing continuously. 
The particles are being swept up by the interfaces as the liquids separate. As the interfacial area decreases the colloids become jammed together, ultimately arresting the phase separation as predicted~\cite{stratford05}. These results are the first demonstration that the route to bijel formation is a slow change in temperature at critical composition.

\begin{figure}[htp]
\centering
\includegraphics[width=1\columnwidth]{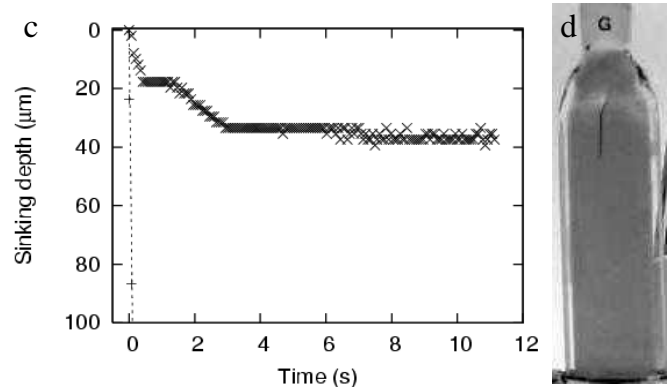}
\caption{{\bf Structure and mechanical properties of bijels.} a: Fluorescence confocal microscopy images at different depths into the sample of 2,6-lutidine-water at critical composition with $\Phi_{v}$=2\%, quenched from room temperature to 40$^{\circ}$C at 17$^\circ$C/min. Depth into sample is shown on labels. The equal darkness of both domains shows that there are few residual particles in either domain.  b: Reconstructions along the vertical axis (thinnest dimension) for bottom, centre and top of images on the left reaching 500~$\mu$m into the cuvette. In both cases scale bar is 100~$\mu$m. Actual sample thickness is 1~mm. c: Sinking depth of cylinder with mass of 1.9~mg and 0.2~mm diameter falling over time in both droplet emulsion (dashed line) and bicontinuous emulsion ($\times$) with compositions as above. The cylinder falls quickly through the droplet emulsion, and much more slowly through the bicontinuous sample with plateaus of zero sedimentation rate. d: Cylinder 12~sec after being released into a bicontinuous emulsion (cuvette is 1~cm wide). The wire remains at this location for weeks confirming that there is a yield stress.}\label{fig:3d_steps}
\end{figure}

We are now able to create three-dimensional bijels routinely using this protocol.  Fig.~\ref{fig:3d_steps}a and \ref{fig:3d_steps}b present the results of a fluorescence confocal microscopy study through 1~mm depth of sample (see Methods). The sample was created via a deep quench to 40$^{\circ}$C at 17$^{\circ}$C/min at critical composition with $\Phi_v$=2\% neutrally wetting silica colloids. After the quench we find an arrested, bicontinuous pattern characterised by a domain size $\xi \approx$ 40~$\mu$m independent of the depth into the sample (see also Supplementary Movie).
Fig.~\ref{fig:3d_steps}b shows vertical reconstructions through the bottom, centre and top of the images seen on the left. Many domains are encountered in traversing the sample along the thinnest dimension (the vertical). 
This is fundamentally different to the two-dimensional structures obtained from phase-separating polymers~\cite{chung05} and fast quenched alcohol-alkane mixtures~\cite{uslangmuir}. In these cases the domain size is comparable to the thickness of the sample making it impossible to distinguish between structural stability and surface effects. We achieve fully three-dimensional samples by quenching through the critical point and exploiting critical slowing down. We demonstrate that spinodal decomposition takes place throughout the depth of the sample (see Supplementary Movie) resulting, in the presence of particles, in a rigid, fully three-dimensional bicontinuous structure. There are clear variations in the mean curvature of the interfaces (on a length scale large compared to the colloids) throughout the static sample suggesting a semi-solid character. The sample is kept at 40$^{\circ}$C in an incubator for several months and scanned periodically to determine whether the structure is stable. Our survey of the morphology of the domains over this time revealed it to be long lived but does not rule out extremely slow ageing of the particle organisation within the walls.

A comparison of the images in Fig.~\ref{fig:spindec} and  Fig.~\ref{fig:3d_steps}a and \ref{fig:3d_steps}b shows clear differences in the colloid populations of the two fluid domains. In Fig.~\ref{fig:spindec} the colloids have a marked preference for one of the fluid phases. By adjusting the colloid drying protocol it is possible to tune the surface chemistry (wettability; see Methods) such that there is little excess of particles in either phase (see Fig.~\ref{fig:3d_steps}a and \ref{fig:3d_steps}b). To successfully create a bijel the appropriate wetting conditions must be obtained for each fluid-fluid-colloid system.

\begin{figure}[htp]
\centering
\includegraphics[width=1\columnwidth]{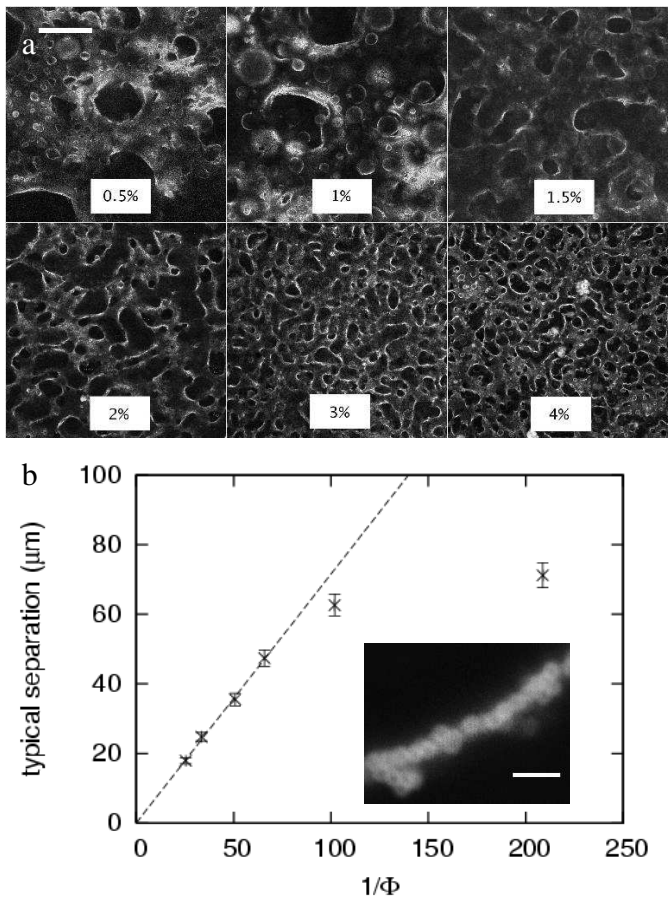}
\caption{{\bf Influence of particle volume fraction on domain size.} a: Fluorescence confocal microscopy images of 2,6-lutidine-water samples at critical composition, quenched from room temperature to 40$^{\circ}$C at 17$^\circ$C/min with varying particle volume fractions $\Phi_{v}$ (scale bar 100~$\mu$m). Particles are shown in white while liquids appear dark. b: Typical separation between domains (see Methods) plotted against inverse of particle volume fraction. Inset: High resolution image of colloid packing, scale bar 1~$\mu$m.}\label{fig:volfrcomp}
\end{figure}

Our protocol allows full control of the interface separation, so that the bijel can be tuned to suit the needs of varied applications. We illustrate this using critical composition 2,6-lutidine-water samples with volume fractions of silica particles between 0.5\% to 4.0\%  warmed to 40$^{\circ}$C at 17$^{\circ}$C/min. The interface separation is extracted from the confocal microscopy images via the calculation of a structure factor (see Methods). Qualitatively, the more particles employed the smaller the characteristic interface separation (Fig.~\ref{fig:volfrcomp}a). At higher volume  fractions of colloids the change in interface separation occurs without a change in morphology and perfect control is possible in this regime. Fig.~\ref{fig:volfrcomp}b presents this quantitatively: the variation in interface separation $\xi$ with $\frac{1}{\Phi_v}$ is linear down to $\Phi_v$=1\% ($\frac{1}{\Phi_v}$=100). The scaling behaviour results from a specific quantity of interface being arrested by the jamming of a specific volume fraction of particles. Experimentally the slope of the linear part of the graph is 0.72~$\pm$~0.02~$\mu$m (Fig.~\ref{fig:volfrcomp}b). This value can be compared to expectations for close-packed monolayer coverage of spherical domains. In this case the droplet diameter $\xi = \frac{\pi}{\sqrt{3}}\frac{d}{\Phi_v}$, where d is the colloid diameter. This gives a slope of 1.05~$\mu$m for our particles which is of the same order as the measured value. The bijel domains are not spherical and the colloids are unlikely to be close-packed, however, as the packing fraction falls below the close-packed limit the slope decreases, therefore the discrepancy is not a cause for concern. The rough agreement suggests that the interfaces may be stabilized by a monolayer of colloids: as in the computer simulations~\cite{stratford05}.
Confirmation is provided by high-resolution imaging (inset to Fig.~\ref{fig:volfrcomp}b). Stabilization by a monolayer of colloids demonstrates that the key to structure formation is the interfacial tension and not direct interactions between colloids. This shows that we have overcome the irreproducibility of the structures shown in~\cite{uslangmuir} which were supported by thick colloidal layers. For 0.5\% and 1.0\%  volume fraction samples these interfaces appear to adopt a preferred curvature which undermines emulsion connectivity observed for all other samples. Other, more complex configurations can be obtained by also varying the warming rate (see Supplementary Note).

The results we report can be connected to the known behaviour of colloid-stabilized droplet emulsions as the proportions of dispersed and continuous phases are varied. Unlike emulsions stabilized by amphiphilic surfactants, colloid-stabilized emulsions can undergo a phase inversion from oil-in-water to water-in-oil due to changes in the volume fractions of fluids alone~\cite{binks02surf, kralchevsky05}. Compositions either side of inversion have been characterised~\cite{aveyard03}, however, it has not previously been possible to create an emulsion within the inversion region itself. We illustrate the connection between the bijel and phase inversion by preparing a series of samples with systematically varying liquid compositions (see phase diagram Fig.~\ref{fig:liqucomp}a). The samples contain $\Phi_v$=2\% silica colloids and were heated to 40$^{\circ}$C at 17$^{\circ}$C/min. At the critical composition the convoluted, three-dimensional structure is formed while all other samples form droplet emulsions (Fig.~\ref{fig:liqucomp}b). We find an inversion taking place (see Methods) with the internal phase changing from lutidine-rich to water-rich around the critical point. We conclude that a bijel is a colloid-stabilized emulsion in the inversion region where neither fluid is the internal phase. Clearly the inversion is here also associated with the demixing kinetics. For off-critical quenches the initial demixing happens via the formation of minority phase nuclei which become the emulsion droplets. Further from criticality the volume of the minority phase falls and arrest occurs at smaller droplet size as required by the fixed volume fraction of particles (Fig.~\ref{fig:liqucomp}b).

Previous studies of droplets~\cite{us05, xu05, tsamantakis05}, bubbles~\cite{subramaniam05, subramaniam06} and extended domains~\cite{uslangmuir, chung05, edmond06} have shown that colloid-stabilized fluid-fluid interfaces are static and can support variations of mean curvature on the macroscale indicating that they are at least semi-solid. A Young's modulus~\cite{vella04} has been inferred from studies of interfacial buckling~\cite{aveyard00col, xu05,tsamantakis05}. Where possible, local rearrangements tend to be preferred to large-scale rearrangements because less bare liquid-liquid interface is exposed~\cite{subramaniam06}. The bijel, too, exhibits macro-scale variations of mean curvature in a static liquid-liquid interface. Since the interface extends through the entire sample its semi-solid character makes the bijel a soft solid. To demonstrate the load-bearing properties of the bijel we look at two samples of the same composition as in Fig.~\ref{fig:3d_steps}a but while the particles stabilize extended domains in the bijel the second sample contains individual droplets. In each case a cylindrical wire of 0.2~mm diameter and total mass 1.9~mg is dropped into the cuvette and the path the weight takes is recorded with a digital camera at 15~frames per second. The images are subsequently processed using Vision Assistant~\cite{visionassistant} to track the position of the weight. The droplet emulsion provides little resistance: the weight sinks to the bottom with a speed of 960 mm/s (see Fig.~\ref{fig:3d_steps}c, dashed line). By contrast, the weight in the bicontinuous emulsion sinks much more slowly and becomes stuck after travelling a few millimetres (see Fig.~\ref{fig:3d_steps}c and Fig.~\ref{fig:3d_steps}d) and occasionally sinks further. Over a period of several weeks the weight remains supported against gravity with little further movement. 
While the slowing and stopping of the weight in the bijel suggests viscoelastic properties, the zero-sedimentation rate plateaus in Fig.~\ref{fig:3d_steps}c are clear evidence for a yield stress. The occasional descent from one plateau to another may be due to local rearrangements in the interfaces in response to the stress. Such intermittent behaviour is generic to many soft solids~\cite{cipelletti05}. Confocal images examining the path within the bijel taken by the cylinder show that after the fall of the weight a `healing effect' takes place. The ripped-open channels of the convoluted structure are re-covered with particles so no purely liquid interfaces are exposed creating a self-supported tunnel through the structure. 
The lower bound on the yield stress for this bijel (Fig.~\ref{fig:3d_steps}d) can be estimated as $\sigma_s$~=~(cylinder~weight)~/~(cylinder~area)~ =~600~ Pa which is the same order of magnitude as $\gamma$/d~\cite{vella04}. We conclude that the convoluted, particle-laden interfaces give solidity to the sample.

The bijel is a soft composite that incorporates two intermediate length scales: the interface separation and the colloid size. Consequently elasticity and permeability can be tuned independently. Our route using colloids on the 100s~nm scale and channels on the 10s~$\mu$m scale produces a novel emulsion with a shelf life of many months at least. 
The interfacial layer is semipermeable~\cite{dinsmore02} and has interstitial patches of bare interface where small reagents, soluble in immiscible fluids~\cite{luisi88}, can meet. We are currently exploring the bijels response to counter-flowing fluids with a view to potential applications as a microreaction and solvent extraction medium. 

\section*{Acknowledgements}
We are grateful to M.~Cates, B.~Binks, T.~Horozov, E.~Kim and H.~Vass for productive discussions. Funding was provided by EPSRC Grants EP~/D076986/1 and EP~/E502652/1.

\section*{Materials and Methods}\label{matmeth}

The 2,6-lutidine-water system exhibits a lower critical solution temperature (T$_c$=34.1$^{\circ}$C, x$_{l}$=0.064) and a relatively symmetric phase diagram; the spinodal demixing region is easily accessible and the obtained final phases have roughly equal volumes and densities~\cite{grattoni93, faizullin91-23}. The particles are St\"{o}ber silica fluorescently tagged with FITC~\cite{stoeber68, blaaderen92} with a hydrodynamic radius of 290~nm and a polydispersity of 0.025 as obtained from dynamic light scattering. We found the wettability of the FITC doped silica surface to strongly depend on the absorbed surface water. To obtain neutrally wetting conditions the particles are dried at 70$^\circ$C over night under vacuum. Deviations from this procedure strongly affect the stabilizing properties of the particles.
Although recent experiments point out new aspects of stabilization by solid particles associated with electrostatic charge~\cite{leunissen07} here the particles are trapped by the interfacial tension~\cite{binksbook}.

\subsection*{Sample preparation}
A suspension of silica colloids in the single-fluid phase of 2,6-lutidine-water is obtained by initially dispersing the particles in water (MilliQ, 18~M$\Omega$) using an ultrasound probe (Sonics \& Material), with 20~kHz for 2~minutes at 3-6~W and then adding 2,6-lutidine (Sigma Aldrich, $\ge$ 99\%, used as received). The resulting suspension is placed in a glass cuvette (Optiglass) of 1~mm pathlength. An aluminium block designed to snugly fit the cuvette is preheated to 40$^\circ$C. When the cuvette is loaded this yields the fastest heating rate of 17$^{\circ}$C/min and is the standard protocol. For slower rates, heating and cooling coils surrounding the block adjust the sample temperature, which is accurately controlled by a PID controller (Lakeshore 331) with a type-K thermocouple (Omega). To examine the emulsion type of the droplet emulsions obtained when varying the liquid ratios we check whether sedimentation or creaming of the emulsion droplets occurs. However, for $\Phi_v$=2\% the weight of the shell makes all droplets sediment. Therefore the droplets were re-made using a lower volume fraction of colloids. For the mechanical test on the bijel the sample is prepared with the 17$^\circ$C/min quench while the droplet counterpart is prepared by vigorously shaking the sample during the quench to avoid structure formation.

\subsection*{Sample characterisation}
The quenched samples were studied with laser fluorescence confocal microscopy. A Nikon TE300 was used in conjunction with the Biorad Radiance 2100 scanner operating an Ar-ion laser. Due to the controlled temperature environment we use an extra long working distance $\times$20 PlanFluor Nikon objective with adjustable correction collar. Depending on refractive index matching and particle volume fraction, visualisation at depths of up to 800~$\mu$m into the sample could be achieved.
The yield stress test was recorded with a Nikon Coolpix 7 mega-pixel camera while the sample was held at 40$^\circ$C.

\subsection*{Length scale extraction using FFT analysis}
The structures we create are static analogues of sponge mesophases~\cite{roux92} and the scattering pattern is well known. To analyse the dominant length scales we calculate the structure factor by radially averaging the squared Fourier transform of the  microscopy images. To improve statistics we average structure factors for several different images. A knee in the log-log plot indicates the wave vector corresponding to the characteristic length scale.  We fit the resulting curves to extract the characteristic length which is found at 20\% decay of the curve (a small increment on the log scale). Image manipulations were carried out using the ImageJ software package~\cite{imageJ} and IDL~\cite{IDL}.

\clearpage

\begin{onecolumn}
\begin{figure}[htp]
\centering
\includegraphics[width=1\columnwidth]{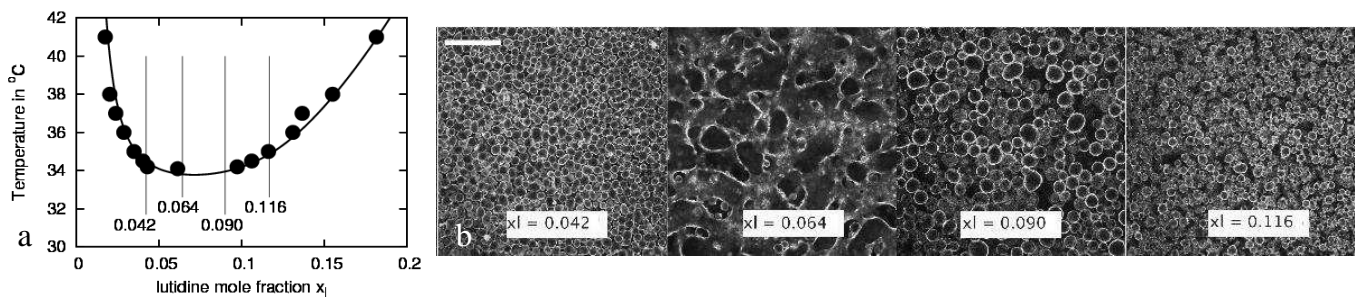}
\caption{{\bf Influence of liquid composition on emulsion morphology.} a: Phase diagram of 2,6-lutidine-water system~\cite{grattoni93}. Vertical lines indicate quench composition corresponding to images on the right. b: Fluorescent confocal microscopy images of water-lutidine samples with $\Phi_{v}$=2\% particles, quenched from room temperature to 40$^{\circ}$C in preheated aluminium block with varying lutidine mole fraction x$_l$ as labelled. A catastrophic phase inversion occurs with the bicontinuous emulsion in the inversion region. Particles are shown in white while liquids appear dark; scale bar 100~$\mu$m.  }\label{fig:liqucomp}
\end{figure}
\end{onecolumn}

\end{document}